# Toroidal Response in Dielectric Metamaterials

**Alexey A. Basharin[1,2], Maria Kafesaki[1,3], Eleftherios N. Economou[1,4] and Costas M. Soukoulis[1,5]**

[1] Institute of Electronic Structure and Laser (IESL), Foundation for Research and Technology Hellas (FORTH), P.O. Box 1385, 71110 Heraklion, Crete, Greece
[2] National Research University "Moscow Power Engineering Institute", 112250 Moscow, Russia
[3] Department of Materials Science and Technology, University of Crete, 71003 Heraklion, Greece
[4] Department of Physics, University of Crete, 71003 Heraklion, Greece
[5] Ames Laboratory-USDOE, and Department of Physics and Astronomy, Iowa State University, Ames, Iowa 50011, USA

*basharin@iesl.forth.gr*

***Abstract* – We present and analyze a dielectric cluster based on subwavelength $LiTaO_3$ polaritonic cylinders for demonstrating toroidal response in THz regime due to mutual coupling of Mie- resonance modes of the cylinders. Based on this cluster, we demonstrate a low-loss metamaterial with the dominant toroidal response, which plays a key role in achieving resonant total transmission.**

## I. INTRODUCTION

Toroidal dipole is not a part of the standard multipole expansion, and was first considered by Zel'dovich to explain parity violation in the weak interaction [1]. Toroidal dipole arises due to the poloidal currents flowing on the surfaces of a torus along its meridians and is characterized by circulating magnetic field confined within the toroid. Toroidal dipole is physically different from electric and magnetic dipoles, as it does not produce electric or magnetic fields in the static case and, while oscillating, scatters electromagnetic radiation weaker than each of the standard dipoles. For this reason, the toroidal response is difficult to be detected on the background of much stronger electric and magnetic moments and thus is usually neglected. However, a special design of subwavelength particles, composing a metamaterial, has enabled to reveal the toroidal dipole mode, due to the suppression of its electric and magnetic multipoles [2]. The structure proposed in [2] exhibited considerable Joule losses, since it was based on conducting elements. Here, the idea of employing dielectric metamaterial clusters for demonstrating toroidal response is for the first time presented. Note that absorption in dielectric materials and metamaterials based on them are considerably smaller than in the case of metamaterials composed of conducting elements.

## II. TOROIDAL RESPONSE DUE TO MIE RESONANCE

In general, to obtain the toroidal dipole excitation one requires to arrange elements with magnetic dipole moments along a circle in a head-to-tail configuration. The toroidal metamolecule in our consideration is constructed from four high-index dielectric subwavelength cylinders, as shown in Fig. 1a. By an external electromagnetic E-polarized field (electric field, E, is parallel to the cylinder axis), circular displacement currents **j** are induced in each of the cylinders (at a specific Mie resonance mode of the cylinders), with opposite directions relative to the axis of the cylinders leading to possibility of high magnetic fields inside them, and thus to magnetic moments **m** directed perpendicularly to the axes of the cylinders, as shown in Fig. 1a. We demonstrate that the circulation of magnetic moments **m** in the dielectric cluster of Fig. 1a can lead to excitation of toroidal moment **T** in the parallel to the cylinder axis direction.

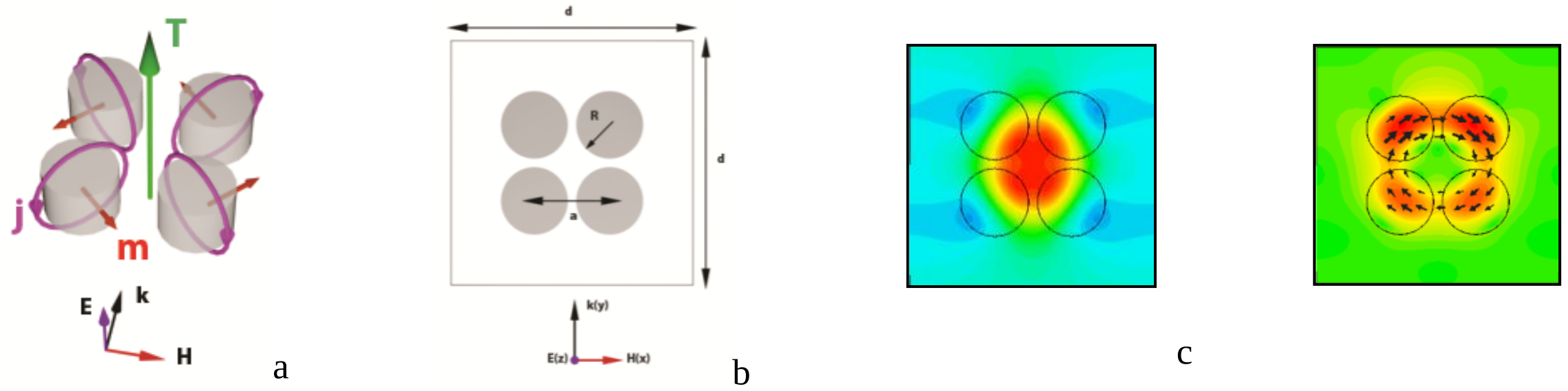

Fig. 1(a): Dielectric metamolecule for toroidal moment excitation based on 4 high-index dielectric cylinders. **j**- displacement current, **m**- magnetic moment, **T**- toroidal moment. (b): Unit cell of the metamaterial consisting of 4 infinitely elongated $LiTaO_3$ cylinders of R=8 μm and the nearest neighbor separation (from center to center) a=18 μm. The square unit cell has a size d=58 μm. (b): Electric field distribution $E_z$ in the metamaterials unit cell at the toroidal resonance mode (at frequency 1.89 THz). (c): Absolute value of magnetic field distribution $H$ at the frequency 1.89 THz.

High-index dielectrics accompanied by low losses occur naturally in microwaves, and are available in the form of ceramics, while in the THz regime, high permittivity and thus high index values can be found in polaritonic materials, just below their phonon-polariton resonance frequency.

The high permittivity values of polaritonic materials (like, e.g. $LiTaO_3$, TlCl, SiC), associated with relatively low losses, allow the observation of a variety of interesting metamaterial phenomena and applications if the high permittivity materials are properly structured. As it will be demonstrated here, polaritonic materials are very promising ingredients for toroidal metamolecules operating in the THz regime [3].

## III. DESIGN OF THE POLARITONIC TOROIDAL METAMATERIAL

We consider a dielectric cluster consisting of four infinitely long cylinders (Fig. 1b) with relative permittivity $\varepsilon_1$ = 41.4, which corresponds to the $LiTaO_3$ polaritonic crystal at frequencies below the phonon-polariton resonance frequency [3] which is around 26.7 THz. In our $LiTaO_3$ cylinders in *vacuum* system the rod radius is $R$=8 μm and the nearest neighbor separation (from center to center) 18 μm (see Fig. 1b). The low-frequency dielectric permittivity of $LiTaO_3$ has a rather high value, static permittivity $\varepsilon_1$=41.4. That certainly makes it stand out among the other polaritonic materials as a candidate for the excitation of Mie-resonant behavior in THz regime. It is also important to note, that at frequencies below 2 THz $LiTaO_3$ posses negligibly low-losses.

We assume that the cylindrical cluster is periodically repeated in ± x direction (embedded in vacuum) with periodicity (lattice constant) d=58 μm (Fig. 1b). The transmission spectra of the system, obtained by the CST Microwave Studio commercial solver, are shown in Fig. 2a. We focus on the third resonance peak (Fig. 2a), which corresponds to the frequency f = 1.89 THz and close to the first-th Mie- resonance mode frequency of the single cylinder, f =2.18 THz. We expect that interference of Mie resonance mode of each cylinder will lead to the emergence of the toroidal response throughout the system. In order to study the nature of the resonance, the radiated powers of multipoles are calculated throught the field distribution in the metamolecule, taking into account current density formalization, and are shown in Fig. 2b [2].

Note that the toroidal response exceeds by several orders the other types of multipolar moments in the vicinity of the frequency f =1.85 THz. In particular, T - moment is more than 5 times the electric moment and 6 times the magnetic moment. This means a significant suppression of other multipoles, suggesting that our structure is an excellent candidate for a metamolecule with dominant toroidal response. It is also important to note that at the frequency f = 1.89 THz the strength of electrical dipole moment is equal to the toroidal one. We argue that destructive interference of their fields outside the metamolecule leads to the cancellation of the structure's net electromagnetic scattering and results in the resonance of total transmission. Furthermore, such combination of toroidal and electric dipolar excitations might correspond to the so-called non-trivial non-radiating charge-current configuration that was predicted to produce waves of vector potential in the absence of electromagnetic fields [4].

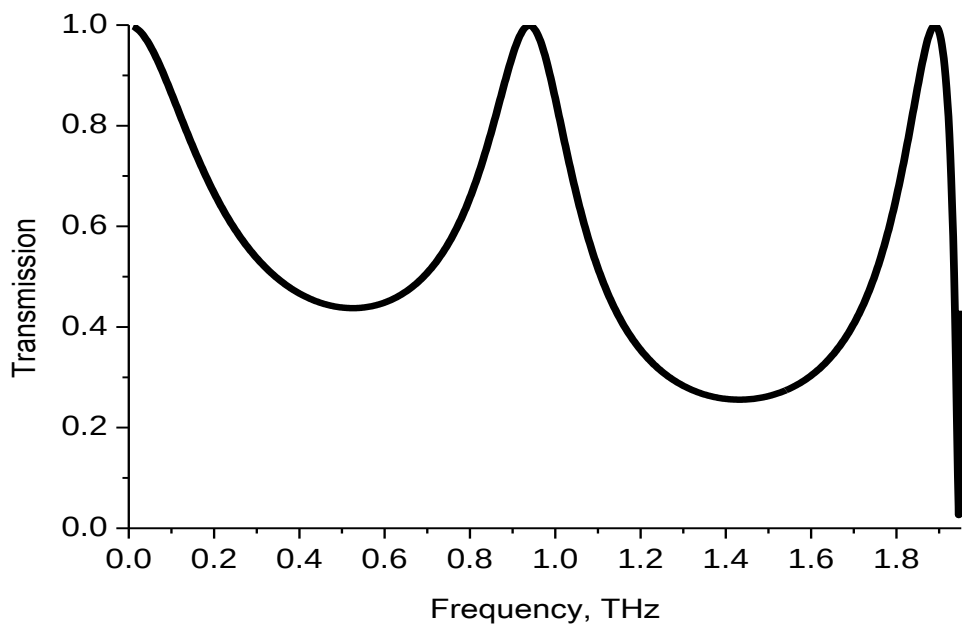

a

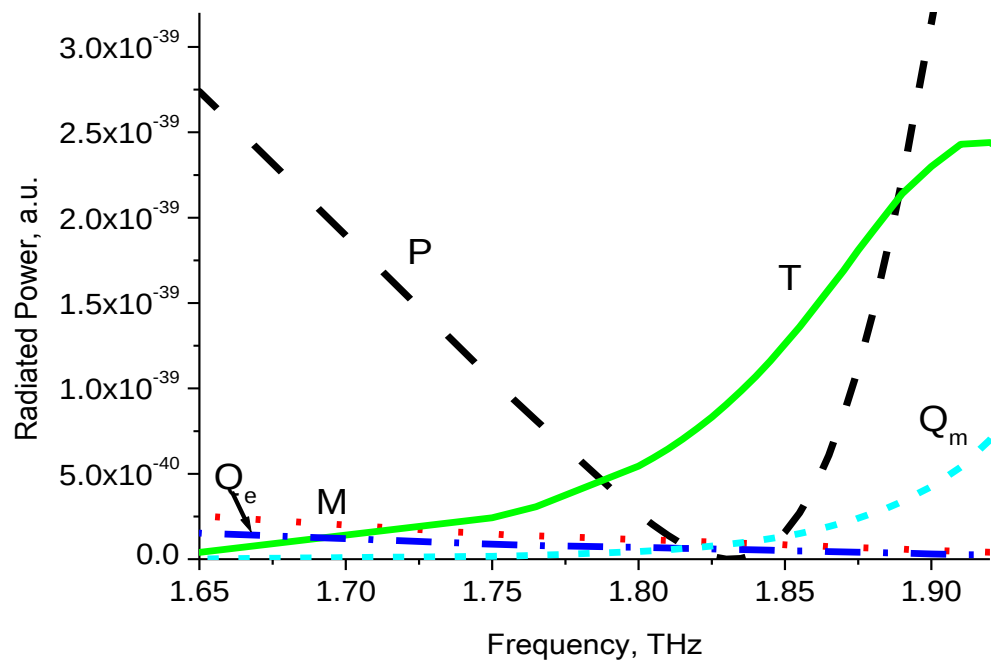

b

Fig. 2(a): Transmission coefficient through the system depicted in Fig. 1b. (b): Radiated power of five multipoles in arbitrary units in the vicinity of frequency 1.82 THz : electric dipole $P$ (black dash curve), magnetic dipole $M$ (red dot curve), toroidal dipole $T$ (green curve), electric quadrupole $Q_e$ (blue dash dot curve), magnetic quadrupole $Q_m$ (cyan short dush curve).

In order to illustrate and explain the resonance corresponding to the toroidal response, we plot the electric and magnetic fields distribution at frequency 1.89 THz (Fig. 1c,d). The incident electromagnetic field excites in each cylinder a Mie-resonance mode characterized by anti-symmetric electric field distribution , as shown in Fig. 1c, which produces a magnetic moment and the circulation of the magnetic field (Fig. 1d) confined within the four-cylinder structure. Such a configuration of closed magnetic field lines indicates the existence of a toroidal dipole mode. Note that the mutual coupling of the modes excited in each cylinder also produces quite strong concentration of the electric field in the central area of the cluster, which is localized on the scale of 1/5λ. Our structure and approach open a new avenue of engineering toroidal response, by exploiting polaritonic metamaterials in THz regime.

## VI. Conclusion

In conclusion we have proposed and theoretically studied clusters of high-index polaritonic cylinders as meta-molecules to achieve toroidal response in the THz regime. The toroidal response is induced by excitation of the Mie-resonance modes in each of the cylinders and by their mutual coupling. The electromagnetic fields of the resulting toroidal mode are strongly concentrated within the structure. To our knowledge this is the first demonstration of toroidal response in dielectric metamaterials, which are characterized by significantly lower dissipative losses compare to toroidal metamaterials made of conducting elements.

## Acknowledgement

Authors acknowledge financial support by Greek GSRT through the project ERC-02 EXEL, and by the EU project ENSEMBLE (grant agreement no. 213669), and the COST Action MP0803. Work at Ames Laboratory was partially supported by the US Department of Energy (Basic Energy Sciences, Division of Materials Sciences and Engineering), contract no. DE-AC02-07CH11358, and by the US Office of Naval Research (award no. N00014-10-1-0925). Work at National Research University "Moscow Power Engineering Institute" was supported by RFBR (grants agreement no. 13-02-00732 and 13-08-01278) and Russian Federal Program "Scientific and scientific-pedagogical Staffs of Innovative Russia" for 2009-2013 (the Agreement no. 14.B37.21.1211). We would like to acknowledge fruitful discussions with Dr. Vassili Fedotov.